\documentclass[conference]{IEEEtran}
\usepackage[numbers]{natbib}
\ifCLASSINFOpdf
\else
\fi
\usepackage{graphicx}
\graphicspath{ {images/} }
\usepackage{multirow}
\usepackage{booktabs}
\usepackage[thinlines]{easytable}
\usepackage[
  separate-uncertainty = true,
  multi-part-units = repeat
]{siunitx}

\begin{document}
%
\title{Single Channel ECG for Obstructive Sleep Apnea Severity Detection using a Deep Learning Approach}



\author{\IEEEauthorblockN{Nannapas Banluesombatkul\IEEEauthorrefmark{1},
Thanawin Rakthanmanon\IEEEauthorrefmark{1}\IEEEauthorrefmark{2}
and
Theerawit Wilaiprasitporn\IEEEauthorrefmark{1}}
\IEEEauthorblockA{
\IEEEauthorrefmark{1}Bio-inspired Robotics and Neural Engineering Lab, \\ School of Information Science and Technology, \\ Vidyasirimedhi Institute of Science \& Technology, Thailand
\\Email: 
nannapas.b@vistec.ac.th, thanawin.r@vistec.ac.th, theerawit.w@vistec.ac.th
}
\IEEEauthorblockA{
\IEEEauthorrefmark{2}Department of Computer Engineering, Kasetsart University, Thailand.
\\Email:
thanawin.r@ku.ac.th}
}
\maketitle


\maketitle 

\begin{abstract}
Obstructive sleep apnea (OSA) is a common sleep disorder caused by abnormal breathing. The severity of OSA can lead to many symptoms such as sudden cardiac death (SCD). Polysomnography (PSG) is a gold standard for OSA diagnosis. It records many signals from the patient's body for at least one whole night and calculates the Apnea-Hypopnea Index (AHI) which is the number of apnea or hypopnea incidences per hour. This value is then used to classify patients into OSA severity levels. However, it has many disadvantages and limitations. Consequently, we proposed a novel methodology of OSA severity classification using a Deep Learning approach. We focused on the classification between normal subjects (AHI $<$ 5) and severe OSA patients (AHI $>$ 30). The 15-second raw ECG records with apnea or hypopnea events were used with a series of one-dimensional Convolutional Neural Networks (1-D CNN) for automatic feature extraction, deep recurrent neural networks with Long Short-Term Memory (LSTM) for temporal information extraction, and fully-connected neural networks (DNN) for feature encoding from a large number of features until it closed to two classes. The main advantages of our proposed method include easier data acquisition, instantaneous OSA severity detection, and effective feature extraction without domain knowledge from expertise. To evaluate our proposed method, 545 subjects of which 364 were normal and 181 were severe OSA patients obtained from the MrOS sleep study (Visit 1) database were used with the k-fold cross-validation technique. The accuracy of 79.45\% for OSA severity classification with sensitivity, specificity, and F-score was achieved. This is significantly higher than the results from the SVM classifier with RR Intervals and ECG derived respiration (EDR) signal feature extraction. The promising result shows that this proposed method is a good start for the detection of OSA severity from a single channel ECG which can be obtained from wearable devices at home and can also be applied to near real-time alerting systems such as before SCD occurs.
\end{abstract}


%
\IEEEpeerreviewmaketitle

\section{Introduction}
Obstructive sleep apnea (OSA) is a common sleep disorder in which complete or partial upper airway obstruction, caused by pharyngeal collapse during sleep \cite{OSA:syndrome}. There are two types of events which can occur during the airway obstruction including hypopnea which is when the inspiratory airflow is reduced, and apnea when it is completely absent for at least ten seconds. This sleep disorder causes loud snoring or choking, frequent awakenings, disrupted sleep, and excessive daytime sleepiness. Furthermore, it is also associated with the incidence and morbidity of hypertension, coronary heart disease, arrhythmia, heart failure, and stroke \cite{OSA:conseq}.

The gold standard of sleep-disordered diagnosis including conditions such as OSA is polysomnography (PSG). It is used to determine the frequency and severity of normal respiratory disorder events per hour and reports as the Apnea-Hypopnea Index (AHI) which can be used to classify the OSA as normal (AHI\textless5), mild (AHI is in 5--14), moderate (AHI is in 15--30), and severe (AHI\textgreater30), respectively \cite{OSA:class}.  However, this method is a form of clinical practice which has to be done overnight in a laboratory or hospital \cite{OSA:treatment} using numerous sensors to acquire the necessary data, such as electroencephalogram (EEG), electrooculogram (EOG), chin electromyography (EMG), leg movement, airflow, cannula flow, respiratory effort, oximetry, body position, electrocardiogram (ECG), and so forth \cite{PSG}.  One study shows that this sleep disorder affects approximately 9-24\% of the general population, and 90\% remain undiagnosed \cite{OSA:undiagnosed} because of the limited number of diagnostic measurements. Since it has to be done in a sleep lab with clinicians, the diagnosis results may be distorted by the lab environment criteria as well as the intrusive and inconvenient measurement sensors which are attached to the patient's body, such as EEG, EOG, chin EMG on the patient's head, and oximetry sensor on the patient's finger.
%
%
%
%

To solve the above issues, there are several studies which tried to diagnose OSA by other methods. \citeauthor{lit:easy_measurements} show that another condition also provides information for the prediction of OSA severity \cite{lit:easy_measurements}. They consequently proposed a new AHI prediction method using only easily available measurements for the estimation of OSA severity level including three BP-related variables, age, BMI, Epworth Sleepiness Scale (a questionnaire), neck circumference, and waistline in order to distinguish the OSA level of normal-mild from moderate-severe with the threshold equal to 15. The results showed that there is a high correlation between the predicted and actual AHI values. However, the accuracy is not reliable enough since the questionnaire results, which are subjective measures, are included in this method. Another solution is trying to detect some OSA-related events to identify the severity of OSA, for example, \citeauthor{lit:snoring} reducing the number of measurement devices attached to the patient's body by placing only the wireless tracheal sensor on the patient's neck and detecting snoring from breathing. Snoring is a significant symptom or sign of OSA \cite{lit:snoring}. Moreover, \citeauthor{lit:bodyposture} also proposed to detect body postures on beds from OSA patients by using 12 capacitively coupled electrodes, and also a conductive textile sheet attached to the patient's chest to acquire electrocardiogram (ECG) signals. Patients with severe OSA have a higher risk of developing cardiovascular diseases, and the apnea occurs more frequently and severely in a supine position than in others \cite{lit:bodyposture}. Nevertheless, most studies still required some specific tools and attached them to several points on the patient's body, which not only uncomfortable for the patient but also leads to some limitations when the patient moves during sleep.

Using wearable devices which provide some necessary bio-sensors for sleep disorder diagnosis such as ECG, EOG, EMG, Heart Rate (HR), Skin Potential and Pulse Oximeter for physiological measurements is obviously a better way because these devices are developed based on the key of unobtrusiveness \cite{lit:wearable}. Moreover, they are also easy to use, easy to find, and cheaper than clinical measurements. One of the most accurate physiological signals provided by various wearable devices and used in several OSA studies are from the ECG \cite{lit:wearableECG1}.

\section{Related Works}

\citeauthor{lit:mhealth} focused on individual differences among all subjects' ECGs and performed a set of IF-THEN rules by a set of parameters related to HRV for each person in order to describe the occurrence of apnea events \cite{lit:mhealth}. Several studies found that ECGs from different people have some identical characteristics and show that the use of only ECG sensors can achieve good accuracy in the detection of sleep apnea. Since various studies show that patients with OSA have slow-wave sleep (SWS) or N3 sleep stage or less than 25\% of their total sleep, \citeauthor{lit:SWS} developed an automatic SWS detection algorithm based on R-R Interval information in both the time domain and frequency domain \cite{lit:SWS}. In another methodology, there are many studies which have used minute-by-minute ECG segments to classify apnea or hypopnea events from normal ECG data \cite{lit:RQA,lit:singlelead,lit:spectrum,lit:hermite,lit:gabor}. 

In 2014, \citeauthor{lit:RQA} proposed an apnea detection method by using recurrence plots (RPs) and subsequent recurrence quantification analysis (RQA) of HRV data which provides the statistical characterization of complex HR regulations. They proposed a combination of classifiers including Support Vector Machine (SVM) and Neural Network (NN) with a soft decision fusion rule for performance improvement \cite{lit:RQA}. In 2015, \citeauthor{lit:singlelead} derived all features from the RR Intervals and ECG derived respiration (EDR) signals to be inputs of classifiers including linear discriminant analysis (LDA), SVM, and least-squares support vector machines (LS-SVM). \cite{lit:singlelead}. \citeauthor{lit:spectrum} also showed that it is possible to acquire information regarding non-stationary signals and their deviation from linearity and Gaussianity using a spectrum and higher order spectrum (HOS) \cite{lit:spectrum}. They consequently developed an algorithm for OSA detection using novel features based on bispectral analysis including a spectrum and HOS of HRV and EDR signals. Afterwards, training with LS-SVM allowed discrimination between normal and apnea episodes. In 2016, \citeauthor{lit:hermite} extracted features based on RR time series from ECG segments along with energy in the error of the QRS approximation and coefficients from the Hermite decomposition. Subsequently, they performed a segment classification between the apnea and normal segments with four classifiers including K-Nearest Neighbor (KNN), Multilayer Perceptron Neural Network (MPNN), SVM, and LS-SVM \cite{lit:hermite}. In 2018, \cite{lit:gabor} decomposed one-minute duration (fixed-length) ECG signals obtained from ECG recordings into band-limited signals using a bank of Gabor filters. After that, they computed phase descriptors (PDs) and extracted histogram features. Finally, they classified apnea events using an LS-SVM classifier with radial basis function (RBF) kernel \cite{lit:gabor}. However, the apnea detection which extracted features from fixed-length segments of ECG signals had some limitations about detailed insights such as the duration of the events because apnea or hypopnea events can occur for less than one minute, or for longer than one minute. 

Instead of fixed-length segmentation, \citeauthor{lit:autoscreen} segmented ECG signals by the iterated cumulative sums of squares (ICSS) algorithm with RR Intervals. They performed feature extractions in the frequency domain and calculated the severity index of each patient. Then, to this value was added some critical subject admission information such as age, body mass index, and gender into an SVM classifier to classify OSA and non-OSA subjects \cite{lit:autoscreen}. Moreover, \citeauthor{lit:HMM} also proposed an OSA detection approach which considers temporal dependence within segmented signals using the Discriminative Hidden Markov Model (HMM). The results showed that the accuracies were improved while classifying OSA events with the use of conventional classifiers with the Discriminative HMM. Nevertheless, the accuracies of all aforementioned works can be affected by the RR Interval detection method because they all used this as a major feature extraction. Furthermore, all of them sought only to detect the apnea or hypopnea events which requires ECG recording for at least one night in order to calculate the AHI value for the purpose of OSA severity detection. \citeauthor{lit:SCD} also showed that the risk of incidence of sudden cardiac death (SCD) was significantly and independently associated with OSA, based both on the frequency of apneas and hypopneas, and the severity of nocturnal oxygen desaturations \cite{lit:SCD}.

Recently, one study \cite{lit:onset} proposed a novel method to detect the severity of OSA without recording ECG for a whole night. They detected the sleep-onset period from the heart rate which is lower at that time than during wakefulness. Then they detected respiration cycles from EDR signals and used this as an input to the AHI prediction using regression analysis. Finally, they used the AHI value to classify the severity of OSA into four levels according to the OSA definition. However, the accuracy of this method is still related to the accuracy of sleep-onset period detection. In consideration of all the aforementioned issues, we aim to initiate an OSA severity detection approach with three main characteristics including: 1) Data acquisition is easier by using only single-channel ECG which can be acquired anywhere from many wearable devices; 2) No overnight sleep study is necessary such that we can prevent the sudden cardiac death (SCD) occurring in severe OSA patients by using only 15 seconds of ECG, and 3) No domain knowledge for feature extraction and feature selection is necessary by using a Deep Learning approach \cite{lit:deep}.

\begin{figure*}
  \includegraphics[width=\textwidth, trim={0cm 9cm 0cm 4cm}]{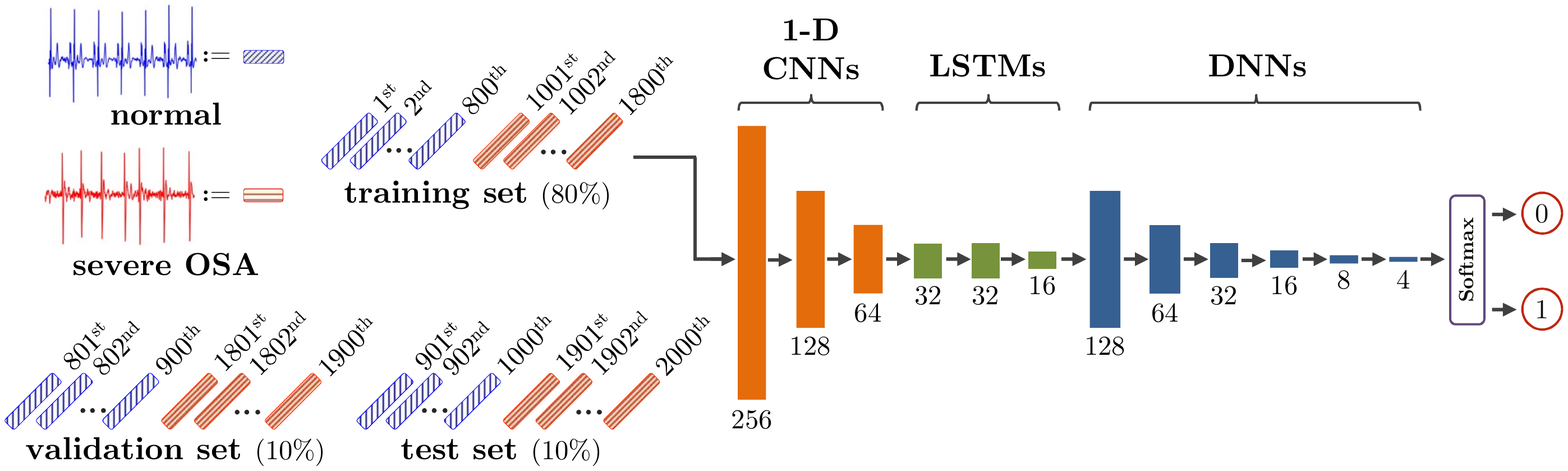}
  \caption{The structure of proposed OSA severity classifier using a Deep Learning approach}
  \label{fig:model}
\end{figure*}

\section{Methodology}

    In this section, we first introduce our data preprocessing procedures. Then, k-fold cross-validation has been applied to test our main classifier. The structure of our classifier will be described at the end of this section.

\subsection{Dataset}
The dataset used in this research was taken from the MrOS sleep study (Visit 1) database \cite{mros1, mros2, mros3, mros4}. The data were recorded with 2911 people of age 65 years or older at 6 clinical centers in a baseline examination. They provide raw polysomnography (PSG) data as European Data Format (EDF) files and annotation XML files of each participant exported from Compumedics Profusion. The ECG signals in this dataset were acquired from Ag/AgCl patch sensors with the sampling rate of 512 Hz through a high-pass filter at 0.15 Hz. Each annotation file includes a marker of start time and duration for the apnea or hypopnea occurrences for each EDF file. We labeled the severity of OSA for each record by using the AHI variable provided in the dataset which is the number of apneas in all desaturations and hypopnea with 4\% desaturation occurring per hour \cite{lit:autoscreen}.  In order to detect severe OSA patients such that we can prevent the sudden cardiac death as mentioned in Section II, we consequently used ECG records from normal patients with AHI values between 2 and 5, and records from extremely severe OSA patients with AHI values greater than 35. Finally, there were 545 subjects including 364 normal subjects and 181 severe OSA subjects which were preprocessed as follows:

\subsubsection{Filter ECGs}
In order to reduce noise in ECG recordings \cite{ECG:denoising}, the original ECG signal was filtered through a notch filter at 60 Hz and then filtered using a bandpass second-order Butterworth filter with cutoff frequencies at 5 and 35 Hz.

\subsubsection{Extract apnea or hypopnea events}
The ECG of each subject was extracted as a sample of apnea or hypopnea events from start time to start time added by duration according to its annotation file. We used only events that lasted 28 - 32 seconds. Then, we selected 30 seconds and normalized it by using the z-score function. Finally, we selected only the first 15 seconds to use.

\subsubsection{Randomly selected samples}
After completing all the data preprocessing procedures, there were 8604 apnea or hypopnea event samples including 3270 samples from normal subjects and 5334 samples from severe OSA patients. We finally randomly selected 1000 samples from each group to use for creating the classifier.

\subsection{K-fold cross-validation}
We used 10-fold cross-validation by separating samples into 3 sets including 80\%, 10\% and 10\% for training, validation, and testing, respectively. Firstly, 1000 samples were selected from each group of patients, and we randomly partitioned them into 10 equal-sized subsamples such that there were 10 subsamples with 100 samples in each. For each subsample, it then remains 900 samples from the total. Then, we randomly selected 100 samples from these to be the validation data and the remaining 800 samples were used as the training data. This process happened 10 times to let all subsamples be tested. Note that the data were combined samples from normal and severe OSA patients together.

\subsection{OSA severity classification using Support Vector Machine}
To compare with our main classifier, we detected RR Intervals and ECG derived respiration (EDR) signals from 15 seconds of ECG samples and extracted features from them which are widely used in several works mentioned in Section II. The set of features included:

\subsubsection{Mean} An average value of the RR intervals.
\subsubsection{Serial correlation} The second and third serial correlation coefficients of the RR intervals.
\subsubsection{pNN50} The number of pairs of adjacent RR intervals where the second RR interval exceeds the first one by more than 50 ms.
\subsubsection{SDSD} The standard deviation of the differences between adjacent RR Intervals.
\subsubsection{Normalized VLF of RR intervals} The normalized power of very low frequency (VLF) of RR Intervals where the total power is the sum of the three components including VLF, low frequency (LF), and high frequency (HF).
\subsubsection{Normalized VLF of EDR} The normalized power of VLF of EDR signal.
\subsubsection{Normalized LF of EDR} The normalized power of LF of EDR signal.
\subsubsection{Ratio of LF to HF of EDR} The ratio of a power of LF to HF of EDR signal.

The range of signal frequency is defined by using 0.003--0.04 as VLF, 0.04--0.15 as LF and 0.15--0.4 as HF.

Then, we applied these features and OSA severity labels as an input of the Support Vector Machine (SVM) model to predict whether each sample was a normal or severe OSA patient.

\subsection{OSA severity classification using a Deep Learning (DL) approach}
As shown in Figure \ref{fig:model}, the training dataset of 1600 samples, 7860 points per sample (15 seconds $\times$ 512 Hz),  is fed into our model. The model is implemented using Keras with parameter configurations as follows:

\begin{itemize}
  \item A stack of one-dimensional Convolutional Neural Networks (1-D CNNs) with 256, 128 and 64 units, respectively, for automatic feature extraction \cite{lit:deep}.
  \item Each CNN layer is followed by batch normalization; the rectified linear unit (ReLU) activation function as well as the max pooling process with pool size equal to 2 in order to extract only important features from the output of its previous layer.
  \item A stack of deep recurrent neural networks with Long Short-Term Memory (LSTM) structure with 128, 128, and 64 units, respectively, with recurrent dropout at 0.4. It is widely used with sequence processing because it is able to preserve the information from the temporal distance between each element in a sequence \cite{model:LSTM}.
  \item Each LSTM layer is followed by a dropout rate at 0.4. 
  \item A stack of fully-connected neural networks (DNNs) with layers of size 128, 64, 32, 16, 8, and 4 hidden nodes for feature encoding from a large number of features into the number closes to 2.
  \item Each DNN layer is followed by the Hyperbolic tangent (tanh) activation function.
  \item The optimizer is \textit{rmsprop} with the learning rate of 0.001.
  \item The \textit{softmax} function is applied for binary classification including normal and severe OSA patients.
\end{itemize}
  
After model optimization, we evaluated our main classifier using accuracy, specificity, sensitivity and F-score. We also compared those metrics with the SVM classifier and performed a paired sample t-test between two classifiers.
  
\begin{figure}
{\includegraphics[width=\linewidth,trim=7.5cm 3.5cm 9cm 3.5cm,clip]{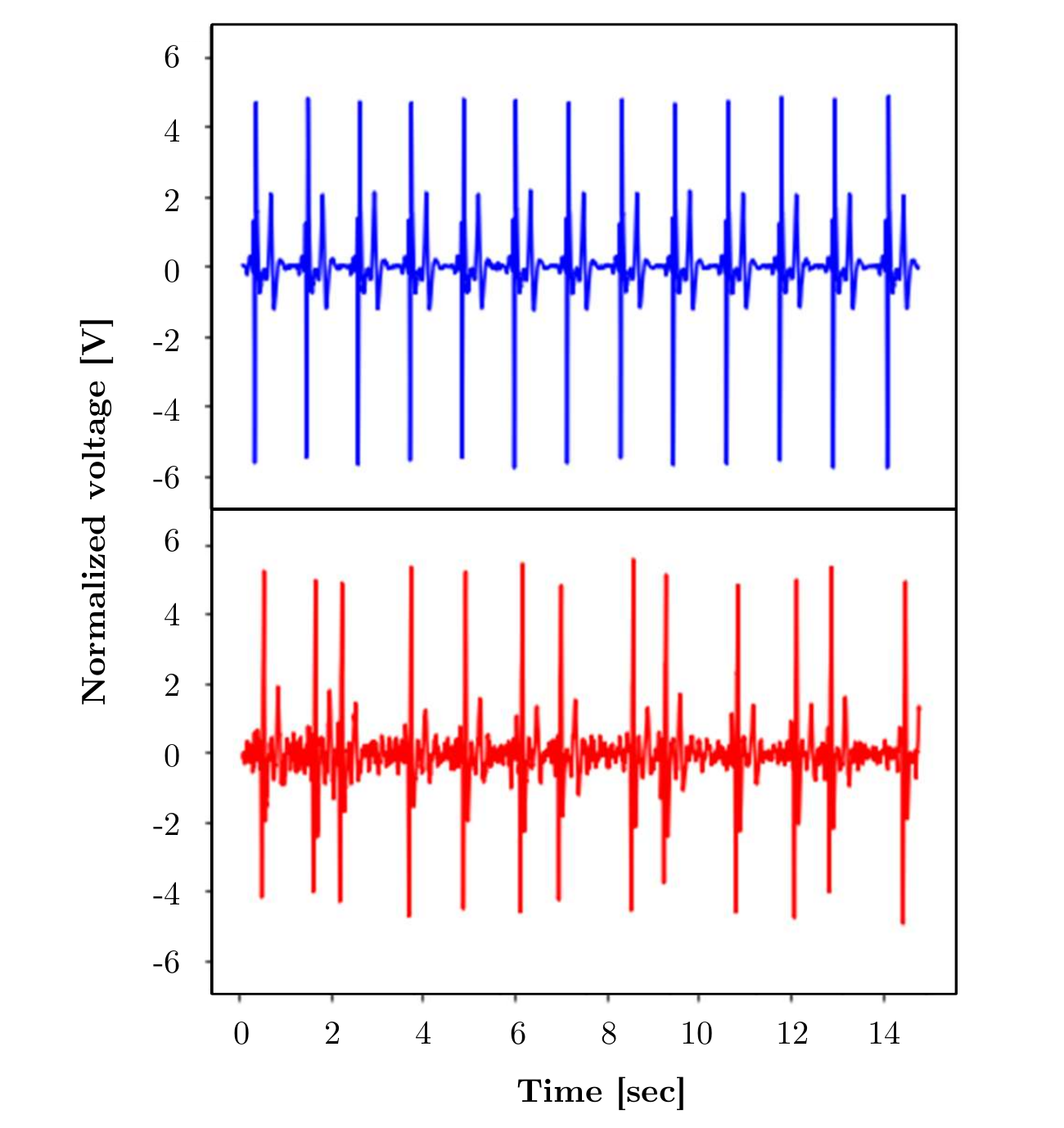}}
  \caption{An example of 15-second normalized ECG signals (512 Hz) with apnea or hypopnea events from a normal patient (top) and a severe OSA patient (bottom)}
  \label{fig:ecg}
\end{figure}

\section{Results and Discussion}
\begin{figure}
  \includegraphics[width=1\linewidth, trim={4cm 3cm 3cm 2cm}]{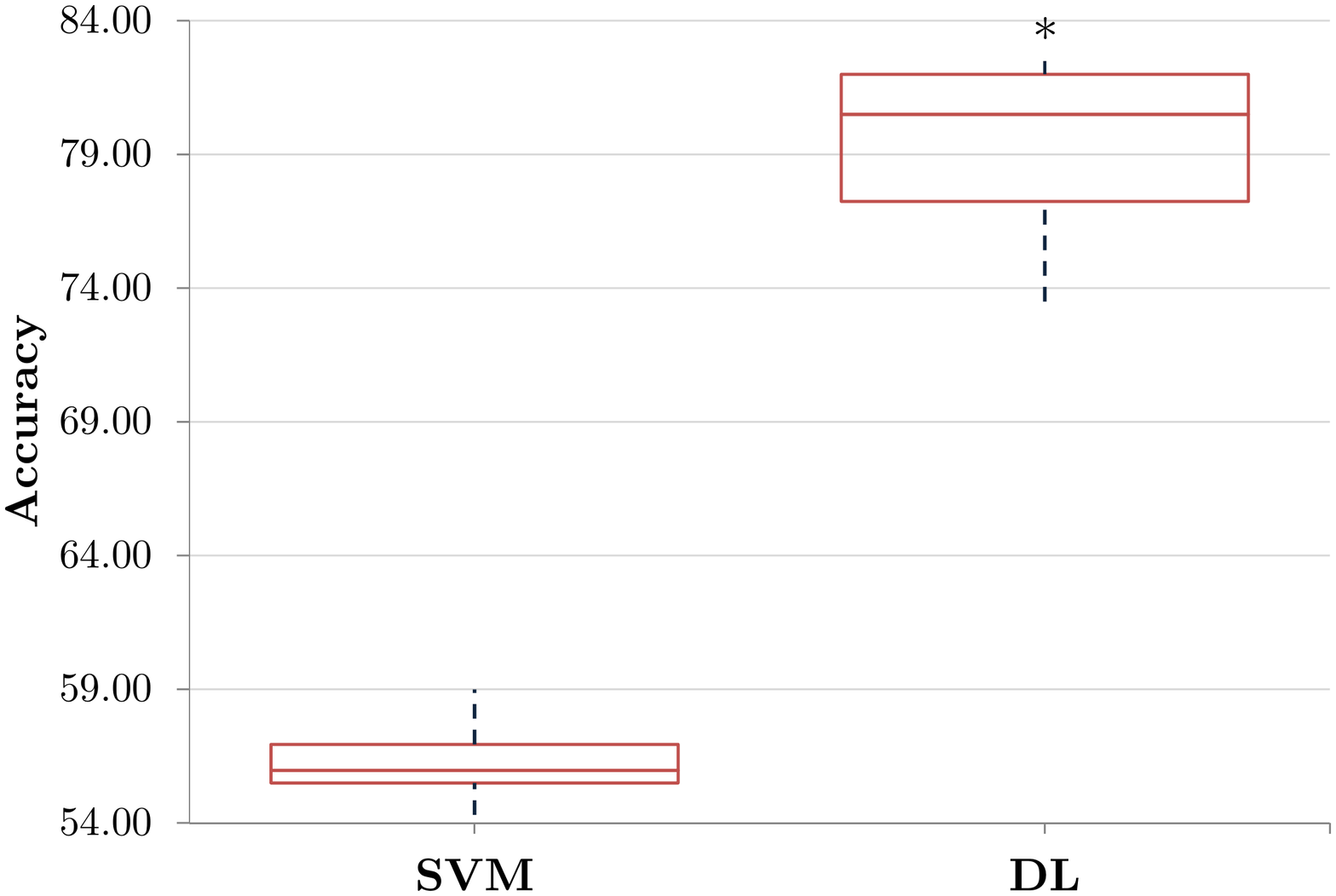}
  \caption{Boxplots of OSA severity classification accuracy in 10-fold cross-validation. * notes that the accuracy of DL approach classifier is significantly higher than the SVM, \emph{p$<$0.05}.}
  \label{fig:Acc}
\end{figure}

An example of 15-second ECG records after normalization from normal and severe OSA subjects is shown in Figure \ref{fig:model}. After training with 10-folds of data in our main classifier with a Deep Learning (DL) approach and the SVM classifier, the results are shown in Table \ref{table:results}. Bold numbers in the table represent the higher value between two classifiers. It shows that our main classifier gives higher accuracy, sensitivity, specificity and F-score in every fold as well as the mean values. Figure \ref{fig:Acc} shows the boxplots of the accuracy acquired from two classifiers. While SVM accuracy ranged from 49.5\% to 59.0\% (mean $\pm$ standard deviation, 55.94\% $\pm$ 2.63\%), our main classifier ranged from 73.5\% to 82.5\% (mean $\pm$ standard deviation, 79.45\% $\pm$ 3.29\%). Consequently, we can conclude that our main classifier performs significantly better than the SVM classifier with \emph{t(10)$=$2.228}, \emph{p\textless0.05}.

Although there are 4 levels of OSA severity including normal, mild, moderate, and severe, as we mentioned previously this paper is only a pilot study of OSA severity classification and focuses on detecting severe OSA patients to prevent sudden cardiac death (SCD), so we started classifying with only 2 classes including normal and severe OSA subjects. As shown in Table \ref{table:results}, the result is promising which means it is a good start to detecting severe OSA patients by using only 15 seconds of ECG signals with apnea or hypopnea events occurring. 

Compared to other previous studies, this proposed classification method achieves a new level of contribution. Firstly, it is more convenient because we use only a single channel of ECG to classify the OSA severity, and such equipment can be found commonly in wearable devices. Secondly, we used only 15 seconds of ECG signals so it is almost instantaneous. From this advantage, it is not only solving a problem of time wasting but also can be adapted to alert the system in order to prevent sudden cardiac death. Lastly, the Deep Learning approach helps us in feature extraction and provides promising results. However, we need to improve the classifier to be compatible with subjects from all OSA severity levels.

\begin{table}[t]
\centering
\caption{Comparison of Accuracies (Acc), Sensitivity, Specificity, and F-Score of normal and severe OSA patients classification between SVM and DL approach in each fold}
\label{table:results}
\begin{tabular}{@{}cp{.4cm}cp{.4cm}cp{.4cm}cp{.4cm}cp{.4cm}cp{.4cm}cp{.4cm}cp{.4cm}cp{.4cm}@{}}
\toprule
\multirow{2}{*}{\textbf{K}}       & \multicolumn{2}{c}{\textbf{Acc.} {[}\%{]}}                            & \multicolumn{2}{c}{\textbf{Sensitivity} {[}\%{]}}                         & \multicolumn{2}{c}{\textbf{Specificity} {[}\%{]}}                                  & \multicolumn{2}{c}{\textbf{F-score} {[}\%{]}}                           \\ \cmidrule(l){2-9} 
                         & {\textbf{SVM}}   & {\textbf{DL}}            & {\textbf{SVM}}   & {\textbf{DL}}             & {\textbf{SVM}}   & {\textbf{DL}}             & {\textbf{SVM}}   & {\textbf{DL}}             \\ \toprule
{1}  & {57.00} & {\textbf{80.50}} & {59.00} & {\textbf{83.00}} & {55.00} 		  & {\textbf{78.00}} & {57.84} & {\textbf{80.98}} \\ \midrule
{2}  & {59.00} & {\textbf{82.50}} & {62.00} & {\textbf{83.00}} & {56.00}          & {\textbf{82.00}} & {60.19} & {\textbf{82.59}} \\ \midrule
{3}  & {58.79} & {\textbf{80.50}} & {48.48} & {\textbf{79.00}} & {69.00}          & {\textbf{82.00}} & {53.93} & {\textbf{80.20}} \\ \midrule
{4}  & {56.78} & {\textbf{82.00}} & {57.58} & {\textbf{77.00}} & {56.00}          & {\textbf{85.00}} & {57.00} & {\textbf{80.21}} \\ \midrule
{5}  & {55.50} & {\textbf{82.00}} & {\textbf{76.00}} & {75.00} & {35.00}          & {\textbf{79.00}} & {63.07} & {\textbf{82.52}} \\ \midrule
{6}  & {55.50} & {\textbf{76.50}} & {55.00} & {\textbf{69.00}} & {56.00}          & {\textbf{84.00}} & {55.28} & {\textbf{74.59}} \\ \midrule
{7}  & {56.28} & {\textbf{82.50}} & {60.61} & {\textbf{88.00}} & {52.00}          & {\textbf{77.00}} & {57.97} & {\textbf{83.41}} \\ \midrule
{8}  & {56.00} & {\textbf{75.00}} & {54.00} & {\textbf{72.00}} & {58.00}          & {\textbf{78.00}} & {55.10} & {\textbf{74.23}} \\ \midrule
{9}  & {49.50} & {\textbf{73.50}} & {61.00} & {\textbf{68.00}} & {38.00}          & {\textbf{79.00}} & {54.71} & {\textbf{71.96}} \\ \midrule
{10} & {55.00} & {\textbf{79.50}} & {67.00} & {\textbf{82.00}} & {43.00}          & {\textbf{77.00}} & {59.82} & {\textbf{80.00}} \\ \midrule \midrule
\textbf{Mean}            & 55.94                      & \textbf{79.45}                      & 60.07                      & \textbf{77.60}                      & 51.80                               & \textbf{80.10}                      & 57.49                      & \textbf{79.07}                      \\ \bottomrule
\end{tabular}
\end{table}

\section{Conclusion and Future Works}
In this paper, the OSA severity classifier using a Deep Learning approach is proposed. We used only 15 seconds of ECG with apnea or hypopnea events from 545 subjects and fed them into a stack of CNNs, LSTMs, and DNNs to classify into 2 classes including normal and severe OSA subjects. We evaluated our proposed classifier by a set of metrics as well as comparison to the SVM classifier with a set of features obtained from RR Intervals and ECG derived respiration (EDR) signal. The proposed classifier is capable of detecting extremely severe OSA subjects from normal subjects with an accuracy level of 79.45\% which is significantly better than the result from the SVM classifier which has an accuracy of 55.94\%. It thus provides an initiation for future development of OSA severity detection systems in order to notify clinicians immediately when severe OSA is found.

Since there are several studies already focused on apnea and hypopnea event detection, we assumed in this paper that we know at which points of ECG that the apnea or hypopnea events  occur. However, the whole system of OSA severity classification should start with detecting apnea and hypopnea onset before using that period to classify the severity. Consequently, the future research study will focus on the methodology of apnea and hypopnea onset detection. Moreover, we will extend the system to support all OSA severity subjects for practical usage.



%

\small 
\bibliographystyle{plainnat}
\bibliography{bibilography.bib}

\end{document}